\documentclass[aps,pra,reprint,superscriptaddress,nofootinbib]{revtex4-2}

\usepackage{amsmath,amssymb,amsthm,mathtools,bm}
\usepackage{braket}
\usepackage{hyperref}
\usepackage{graphicx}
\usepackage{xcolor}
\usepackage{float}
\newcommand{\Tr}{\operatorname{Tr}}
\newcommand{\id}{\mathbb{I}}
\newcommand{\cH}{\mathcal{H}}

\newcommand{\cF}{\mathcal{F}}
\newcommand{\cM}{\mathcal{M}}
\newcommand{\cG}{\mathcal{G}}
\newcommand{\Wplus}{\mathcal{W}_{+}}
\newcommand{\conv}{\operatorname{conv}}
\newcommand{\cl}{\operatorname{cl}}

\newcommand{\Wit}{\operatorname{Wit}}
\newcommand{\cone}{\operatorname{cone}}

\theoremstyle{plain}

\theoremstyle{definition}

\theoremstyle{remark}

\begin{document}
%===========================================================

\title{Activating entanglement and EPR steering from continuous-variable resources using witness-based measures}
\author{Kaustav Chatterjee}
 \email{kauch@dtu.dk}
\author{Ulrik Lund Andersen}%
 \email{ulrik.andersen@fysik.dtu.dk}
\affiliation{%
Center for Macroscopic Quantum States (bigQ), Department of Physics, Technical University of Denmark, \\
 Building 307, Fysikvej, 2800 Kongens Lyngby, Denmark %\textbackslash\textbackslash
}%
\date{\today}

\begin{abstract}
We introduce a general witness-based framework for quantifying and operationally activating continuous-variable (CV) resources into discrete-variable (DV) bipartite entanglement or Einstein-Podolsky-Rosen (EPR) steering. For the three standard CV resource theories associated with Wigner negativity (WN), genuine non-Gaussianity (GNG), and standard non-Gaussianity (SNG), we define infinite families of bounded-witness monotones indexed by box constraints on the witness operators. For closed convex free sets, these monotones are faithful, strongly monotonic under free instruments, Lipschitz continuous, and convex. For closed nonconvex free sets, we show that faithfulness requires a two-copy lift and formulate the corresponding strong-monotonicity statement in the lifted theory. We further construct witness-dependent completely positive trace-preserving (CPTP) measure-and-prepare channels whose outputs are two-qubit Werner states. For the representative case $n=m=1$, the optimal entanglement and EPR steering attainable within this witness-dependent activation family are exactly proportional to the underlying monotones. We illustrate the framework with odd-parity states, pure-loss single-photon states, and Gottesman-Kitaev-Preskill (GKP) states, and derive explicit lower bounds for pure-state GNG and SNG. More broadly, our results show that closed CV free sets admit witness-based quantifiers with a direct operational interpretation in terms of experimentally accessible DV correlations.
\end{abstract}

\maketitle

%===========================================================
\section{Introduction}
%===========================================================
Quantum resource theories \cite{Chitambar_2019,Gour_2025} provide a unified framework for identifying and quantifying the nonclassical features of quantum states that enable advantages \cite{grover1996fastquantummechanicalalgorithm,Shor_1997,huang2025vastworldquantumadvantage} in information processing, communication, sensing, and computation \cite{asera,Braunstein_2005,andersen2010continuousvariablequantuminformation}. In such a framework, one specifies a set of free states together with a class of free operations, and any state outside the free set is regarded as resourceful. Entanglement theory is the canonical example. In continuous-variable (CV) quantum systems, however, additional and physically important notions of resourcefulness arise, reflecting experimentally relevant constraints on state preparation, control, and measurement.

In CV systems, different choices of free states lead to distinct and nested resource theories. Particularly important are the sets of Gaussian states, convex mixtures of Gaussian states, and states with nonnegative Wigner functions. These give rise, respectively, to the notions of standard non-Gaussianity, genuine non-Gaussianity, and Wigner negativity. Such resources play central roles in CV quantum information processing: Gaussian states and Gaussian operations are often viewed as comparatively easy to prepare and manipulate, whereas departures from the Gaussian setting are required for a variety of tasks, including universal quantum computation, quantum error correction, and nonclassical state engineering.

A central problem in any resource theory is to quantify the amount of resource carried by a given quantum state. For CV resources, many of the best-known quantifiers are geometric in nature, being defined through a distance, divergence, or robustness with respect to the relevant free set t\cite{Genoni_2007,Genoni_2008,NG1}. These measures are conceptually natural and often mathematically powerful, but they are not always closely connected to how resources are identified in practice. Experimentally, nonclassicality is often detected not through a global optimization procedure, but through the measurement of suitable observables whose expectation values certify that a state lies outside the free set. This motivates the search for witness-based resource quantifiers that retain the structure of a proper monotone while remaining directly connected to experimentally meaningful observables.

Witness operators are well established in entanglement theory \cite{Brand_o_2005}, where they serve both as practical detection tools and, in suitable forms, as the basis of quantitative resource measures. In contrast, witness-based quantification of CV resources remains much less developed. This is especially relevant in the CV setting, where natural witness observables often already exist, such as displaced parity operators for detecting Wigner negativity. A general witness-based framework is therefore desirable both conceptually and operationally: conceptually, because it places different CV resources within a common geometric language, and operationally, because it ties resource quantification to observables that can, at least in principle, be accessed experimentally.

In this work, we introduce a family of bounded witness-based monotones for general resource theories with a particular focus on CV resources. Given a free set $F$, we consider Hermitian witnesses that are nonnegative on all free states and optimize their negative expectation value over a bounded operator interval. This leads to a family of measures $\cM_{nm}(\rho;\cF)$ that quantify how strongly a state violates the free-state constraints within a prescribed witness class. When the set of free states is convex, we show that these quantities satisfy the key properties expected of a resource monotone: They are faithful, strongly monotonic under free instruments, invariant under free unitary processes, and Lipschitz continuous. For closed nonconvex free sets, such as the set of Gaussian states, we show that a faithful single-copy linear witness theory is obstructed by convexification; the correct formulation instead lives on a lifted two-copy free set, where faithfulness and strong monotonicity can again be established. We further show, however, that strong monotonicity is not guaranteed on the single copy level, even though monotonicity is. For the representative case $n=m=1$, the construction also admits a natural dual interpretation in terms of the trace-norm distance to the cone generated by the free states.

Our main result is that these witness monotones admit an exact operational interpretation through activation into standard discrete-variable quantum correlations. Specifically, for every feasible witness we construct a measure-and-prepare quantum channel whose output is a two-qubit Werner state. The entanglement or EPR steering of this output state is determined directly by the expectation value of the chosen witness on the input CV state. Optimizing over the witness family then shows that the maximal activated entanglement or steering is exactly proportional to the corresponding witness monotone. In this way, the resource content of a CV state is not merely bounded by an operational task, but can be faithfully encoded into experimentally accessible discrete-variable signatures.

We then specialize the framework to Wigner negativity, which provides the clearest and most physically transparent realization of the general construction. In this case, displaced parity operators form a natural witness family, and the resulting monotone is directly related to the negativity depth of the Wigner function. This yields a simple operational picture: Negative regions of the Wigner function can be mapped into bipartite entanglement or EPR steering of a two-qubit output state through an explicit activation protocol. We illustrate the framework with several examples, including odd-parity states and single photons subjected to pure loss, for which the activation behavior exhibits a sharp threshold. In our examples, we choose GKP states as non-Gaussian states and separate an analytic monotonicity argument based on Gaussian displacement noise from a numerical finite-energy study based on damped GKP codewords undergoing loss and one round of error correction. We also derive explicit lower bounds for pure-state GNG and SNG, giving operational lower bounds on the entanglement and steering that can be activated from such states.

The paper is organized as follows. In Sec.~\ref{sec:prelim} we review the CV free sets and the two-qubit Werner family used in the activation protocol. In Sec.~\ref{sec:fam} we introduce the witness-based monotones, first for closed convex free sets and then for closed nonconvex free sets via the two-copy lift. Section~\ref{sec:actt} presents the entanglement and EPR-steering activation channels. In Sec.~\ref{sec:examples} we discuss explicit examples, including odd-parity states, pure-loss single photons, GKP states, and pure-state lower bounds for GNG and SNG.
%===========================================================
\section{Preliminaries}
\label{sec:prelim}
%===========================================================
Throughout this paper, $\cH$ denotes a separable Hilbert space, finite or infinite dimensional as appropriate from context. We write $B(\cH)$ for the bounded operators on $\cH$, $T(\cH)$ for the trace-class operators, and $D(\cH)$ for the density operators. When discussing closedness in infinite dimension we use the trace-norm topology on $T(\cH)$ \cite{Turner_2025}, unless explicitly stated otherwise. In particular, $D(\cH)\subseteq T(\cH)\subseteq B(\cH)$.
\subsection{Different CV resources}
Quantum resource theories\cite{Gour_2025,Chitambar_2019} formalize the idea that any realistic agent is restricted in which states can be prepared and which operations can be implemented at negligible cost. One therefore singles out a set of free states $\cF\subseteq D(\cH)$ and studies the complementary resourceful states $D(\cH)\setminus \cF$. The corresponding free operations are CPTP maps $\Lambda$ satisfying $\Lambda(\cF)\subseteq \cF$. Throughout this work we assume that the free set is trace-norm closed. Whether it is convex or not will play an essential role later.

We focus first on single-mode CV systems, although all constructions extend directly to the multimode setting. In phase-space language the system is described by quadrature operators $\hat r=(\hat x,\hat p)$ satisfying $[\hat x,\hat p]=i$. Equivalently one may use annihilation and creation operators $\hat a,\hat a^\dagger$ with $[\hat a,\hat a^\dagger]=1$. States may be represented either in the quadrature basis or in the Fock basis $\{\ket{n}\}_{n\ge 0}$. The parity operator is
\begin{align}
    \hat\Pi=\sum_{n=0}^\infty (-1)^n\ket{n}\bra{n},
\end{align}
and its displaced version is $\hat\Pi(\alpha)=\hat D(\alpha)\hat\Pi\hat D(-\alpha)$, where $\hat D(\alpha)=\exp(\alpha\hat a^\dagger-\alpha^*\hat a)$ is the Weyl displacement operator for $\alpha\in\mathbb C$.

A state $\rho$ admits an equivalent phase-space description through its Wigner function\cite{asera,Braunstein_2005}
\begin{align}
    W_\rho(\alpha)=\frac{2}{\pi}\Tr[\hat\Pi(\alpha)\rho].
\end{align}
The Wigner function is a real-valued quasiprobability distribution and may therefore take negative values. Its marginals, however, are ordinary probability distributions and are directly accessible, for example, by homodyne detection. The three free-state sets relevant for this work are the following:
\begin{itemize}
    \item $\Wplus:=\{\rho\in D(\cH): W_\rho(\alpha)\ge 0\ \forall\alpha\in\mathbb C\}$. This is the set of Wigner-positive states; it is closed and convex.
    \item $\cG$, the set of Gaussian states. These are precisely the states with Gaussian Wigner functions. In the single-mode case they are completely specified by the first moments $\mu=\Tr(\hat r\rho)$ and covariance matrix $V_{jk}=\Tr\!\big(\rho\{\hat r_j-\mu_j,\hat r_k-\mu_k\}\big)$, which obeys the bona fide condition $V+i\Omega\ge 0$ with
    \begin{align}
        \Omega=\begin{pmatrix}
        0&1\\
        -1&0
    \end{pmatrix}.
    \end{align}
    Their Wigner function can be written explicitly as
    \begin{align}
        W_\rho(r)=\frac{1}{\pi\sqrt{\det V}}\exp\!\big[-(r-\mu)^T V^{-1}(r-\mu)\big],
    \end{align}
    where $r=(x,p)$. The set $\cG$ is closed but not convex.
    \item $\bar{\cG}_c:=\cl(\conv(\cG))$, the trace-norm closed convex hull of Gaussian states. This is the free set for genuine non-Gaussianity and is closed and convex by construction.
\end{itemize}
 A canonical hierarchy of state sets is \cite{Takagi_2018}
\begin{align}
\cG \subset \bar{\cG}_c \subset \Wplus,
\label{eq:cv-hierarchy}
\end{align}
In what follows we use $\cF$ as a placeholder for the free set under consideration. Choosing $\cF=\Wplus$ gives the WN theory, choosing $\cF=\bar\cG_c$ gives the GNG theory, and choosing $\cF=\cG$ gives the SNG theory.
\subsection{Quantum information properties of two qubit Werner states}
We shall repeatedly use the two-qubit Werner family, so we briefly collect the relevant formulas. Throughout this subsection we work on $\cH\otimes\cH$ with $\dim \cH=2$. The Werner states are
\begin{align}
    w(q)=q\ket{\psi^-}\bra{\psi^-}+\frac{1-q}{4}\id,\qquad q\in\left[-\frac{1}{3},1\right],
\end{align}
where $\ket{\psi^-}=(\ket{01}-\ket{10})/\sqrt 2$ is the singlet. These states are characterized by their invariance under collective local unitaries,
\begin{align}
    w(q)=(U\otimes U)w(q)(U\otimes U)^\dagger.
\end{align}
We will use the following standard properties of the Werner family\cite{LI2021168371}:
\begin{itemize}
    \item \emph{Entanglement.} For two qubits, the negativity is an easily computable faithful entanglement monotone. On Werner states,
    \begin{align}
        E(w(q))=
        \begin{cases}
            \dfrac{3q-1}{4}, & q\in\left(\dfrac{1}{3},1\right],\\[1ex]
            0, & q\in\left[-\dfrac{1}{3},\dfrac{1}{3}\right].
        \end{cases}
    \end{align}
    It is often convenient to reparametrize the family via $q=(4p-1)/3$, which gives
    \begin{align}
        w(p)=p\ket{\psi^-}\bra{\psi^-}+(1-p)\frac{\id-\ket{\psi^-}\bra{\psi^-}}{3}.
    \end{align}
    This form makes explicit the decomposition into the singlet and the maximally mixed state on the triplet subspace.

    \item \emph{Geometric discord.} Given a local von Neumann measurement $P^A=\{P_k^A\}_k$ on subsystem $A$, the post-measurement classical--quantum state is
    \begin{align}
        P^A(\rho)=\sum_k (P_k^A\otimes \id)\rho(P_k^A\otimes \id).
    \end{align}
    In the Hilbert--Schmidt convention used in Ref.~\cite{LI2021168371}, the geometric discord is
    \begin{align}
        D(\rho)=\min_{P^A}\|\rho-P^A(\rho)\|^2,
    \end{align}
    and for Werner states one has $D(w(q))=q^2/2$.

    \item \emph{EPR steering.} Steering lies strictly between entanglement and Bell nonlocality. Given a POVM assemblage $\{K_{a|x}\}_{a,x}$ on subsystem $A$, the induced subnormalized states on $B$ are
    \begin{align}
        \sigma_{a|x}=\Tr_A\big[(K_{a|x}\otimes \id)\rho\big].
    \end{align}
    The assemblage is unsteerable if there exists a local-hidden-state model, i.e., states $\sigma_\lambda$ and conditional probabilities $p(a|x,\lambda)$ such that
    \begin{align}
        \sigma_{a|x}=\int p(a|x,\lambda)\,\sigma_\lambda\,d\mu(\lambda)
        \qquad \forall a,x.
    \end{align}
    Werner states are steerable if and only if $q>1/2$. This motivates the one-parameter steering monotone
    \begin{align}
        S(w(q))=\max\{0,2q-1\}.
    \end{align}

    \item \emph{CHSH violation.} A convenient measure of maximal CHSH violation strength within the Werner family is
    \begin{align}
        N(w(q))=\max\{0,2\sqrt 2\,q-2\}.
    \end{align}
\end{itemize}

\begin{table}[t]
\centering
\caption{Hierarchy of quantum correlations for the two-qubit Werner state $w(q)=q\ket{\psi^-}\bra{\psi^-}+(1-q)\mathbb I/4$.}
\label{tab:werner_hierarchy}
\begin{tabular}{c|c}
\hline
Range of $q$ & Correlation property of $w(q)$ \\
\hline
$-\frac{1}{3} \le q \le \frac{1}{3}$ & Separable (hence unsteerable and Bell local) \\[0.5ex]
$\frac{1}{3} < q \le \frac{1}{2}$ & Entangled but unsteerable \\[0.5ex]
$\frac{1}{2} < q \le \frac{1}{\sqrt 2}$ & Steerable but CHSH-local \\[0.5ex]
$\frac{1}{\sqrt 2} < q \le 1$ & Violates CHSH, hence Bell nonlocal \\
\hline
\end{tabular}
\end{table}

\section{Family of measures}\label{sec:fam}
In full generality, let $\cF$ be a designated set of free states and let $\rho\notin \cF$ be the state whose resource content we want to quantify. The minimal standing assumption throughout is that $\cF$ is trace-norm closed. Convexity, however, fundamentally changes the structure of witness-based quantification, so we discuss the convex and nonconvex cases separately.
\begin{figure}[H]
    \centering
    \includegraphics[width=0.6\linewidth]{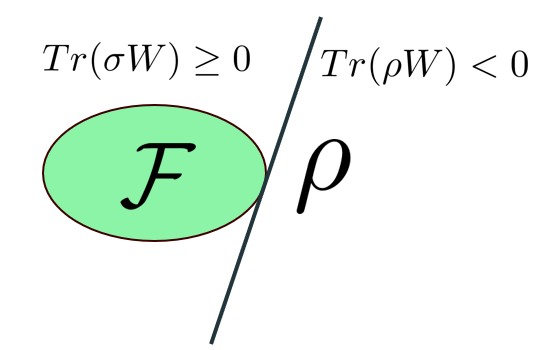}
    \caption{A witness for a resourceful state $\rho$ relative to a free set $\cF$ is a bounded Hermitian operator $W$ such that $\Tr(W\sigma)\ge 0$ for all $\sigma\in\cF$, while $\Tr(W\rho)<0$.}
    \label{fig:witness_geometry}
\end{figure}
\subsection{Convex $\cF$}\label{fam1}
Assume first that $\cF$ is closed and convex. Then for every $\rho\notin \cF$ there exists a bounded Hermitian operator $W\in B(\cH)$ such that $\Tr(W\sigma)\ge 0$ for all $\sigma\in \cF$ and $\Tr(W\rho)<0$. This follows from Hahn--Banach separation applied to the real vector space of Hermitian trace-class operators\cite{Turner_2025,Wang_2018}. We therefore define
\begin{align}
    \Wit(\cF):=\{W=W^\dagger\in B(\cH):\Tr(W\sigma)\ge 0\ \forall\sigma\in \cF\}
\end{align}
and the witness box
\begin{align}
    K_{nm}:=\{W=W^\dagger\in B(\cH):-n\id\le W\le m\id\},  n,m>0.
\end{align}
The bounded-witness monotones are then defined by
\begin{align}
    \cM_{nm}(\rho;\cF):=\sup_{W\in \Wit(\cF)\cap K_{nm}}[-\Tr(W\rho)]_+.
\end{align}
Here $[x]_+:=\max\{x,0\}$. In finite dimension---for example after a Fock-space cutoff---the feasible set is compact and the supremum may be replaced by a maximum. Also it is worth stressing that the use of positive part is not necessary and serves as a reminder that the measure is non-negative. To see this one can just consider the $0$ operator which is feasible always and then the lowest value the measure takes is $0$. These monotones enjoy the following basic properties.\\

\emph{(1) Faithfulness:} The monotone is faithful in the sense that $\cM_{nm}(\rho;\cF)=0\iff \rho\in \cF$.
    \begin{proof}
If $\rho\in \cF$, then $\Tr(W\rho)\ge 0$ for all $W\in \Wit(\cF)$, hence the supremum equals $0$.
Conversely, if $\rho\notin \cF$, the separating hyperplane theorem yields a continuous functional $\ell$ such that
$\sup_{\sigma\in \cF}\ell(\sigma)<\ell(\rho)$ which using dual space representation can be written as $\ell(X)=\Tr(AX)$ for bounded Hermitian $A$.
Let $\alpha:=\sup_{\sigma\in \cF}\Tr(A\sigma)$ and set $W_0:=\alpha\id-A$.
Then $W_0\in \Wit(\cF)$ and $\Tr(W_0\rho)<0$. Since $W_0$ is bounded, scaling
$W:=tW_0$ with $t=\min\{n,m\}/\|W_0\|_\infty$ enforces $W\in K_{nm}$ while preserving $\Tr(W\rho)<0$,
implying $\cM_{nm}(\rho;\cF)>0$.
\end{proof}
    
\emph{(2) Strong Monotonicity:} A completely positive instrument $\{\Lambda_k\}_k$ is called \emph{$\cF$-free} if
each $\Lambda_k$ is trace--nonincreasing, $\sum_k \Lambda_k$ is trace-preserving,
and for all $\sigma\in \cF$,
\begin{equation}\label{eq:free_instrument_cone}
\Lambda_k(\sigma)\in \cone(\cF)\quad(\forall k),
\end{equation}
i.e.\ $\Lambda_k(\sigma)=p_k \sigma_k$ with $p_k\ge 0$ and $\sigma_k\in \cF$. Here $\cone(\cF)$ is set of un-normalized free states. For any state $\rho$, define $p_k:=\Tr[\Lambda_k(\rho)]$ and $\rho_k:=\Lambda_k(\rho)/p_k$ (if $p_k>0$).
Then
\begin{equation}\label{eq:strong_mono}
\cM_{nm}(\rho;\cF)\ \ge\ \sum_k p_k\,\cM_{nm}(\rho_k;\cF).
\end{equation}
\begin{proof}
    Fix feasible $W_k\in \Wit(\cF)\bigcap K_{nm}$ and set $W:=\sum_k \Lambda_k^\dagger(W_k)$. Then we have \emph{(Witness preservation):} For any $\sigma\in \cF$, $\Lambda_k(\sigma)\in\cone(\cF)$ by~\eqref{eq:free_instrument_cone}, so $\Tr(\Lambda_k^\dagger(W_k)\sigma)=\Tr(W_k\Lambda_k(\sigma))\ge 0$; summing gives $W\in\Wit(\cF)$. \emph{(Box preservation):} Since $-nI\le W_k\le mI$, positivity of $\Lambda_k^\dagger$ implies
$-n\,\Lambda_k^\dagger(I)\le \Lambda_k^\dagger(W_k)\le m\,\Lambda_k^\dagger(I)$.
Summing and using $\sum_k \Lambda_k^\dagger(I)=I$ (because $\sum_k\Lambda_k$ is trace-preserving) yields
$-nI\le W\le mI$. Finally,
\[
-\Tr(W\rho)=\sum_k -\Tr(W_k\Lambda_k(\rho))=\sum_k p_k\bigl(-\Tr(W_k\rho_k)\bigr).
\]
Taking the supremum over $W$ on the left and over each $W_k$ on the right gives~\eqref{eq:strong_mono}.
\end{proof}
For unitary free operations $U_F$ that map free states to free states it follows that these measures are invariant meaning $\cM_{nm}(\rho;\cF)=\cM_{nm}(U_F\rho U_F^\dagger,\cF)$. This follows from the fact that for any witness $W$ in the feasible set, $U_FWU_F^\dagger$ is also in the feasible set.\\
\\
\emph{(3) Convexity:}
For any $\cF$ and any $n,m>0$, the map $\rho\mapsto \cM_{nm}(\rho;\cF)$ is convex:
\begin{equation}
\cM_{nm}(p\rho+(1-p)\sigma;\cF)\ \le\ p\cM_{nm}(\rho;\cF)+(1-p)\cM_{nm}(\sigma;\cF)
\end{equation}
with $p\in [0,1]$. 
\begin{proof}
    For fixed $W$, the map $\rho\mapsto -\Tr(W\rho)$ is affine, and $\rho\mapsto[-\Tr(W\rho)]_+=\max\{0,-\Tr(W\rho)\}$ is convex. The pointwise supremum over all feasible $W$ is therefore convex.
\end{proof}

\emph{(4) Lipschitz continuity:} Set $c:=\max\{n,m\}$.  Then for all states $\rho,\sigma$,
\begin{equation}\label{eq:lipschitz_single}
\big|\cM_{nm}(\rho;\cF)-\cM_{nm}(\sigma;\cF)\big|
\ \le\ c\,\|\rho-\sigma\|_1.
\end{equation}
\begin{proof}
Let $c:=\max\{n,m\}$. For any feasible $W$ we have $\|W\|_\infty\le c$, hence
$
\big|-\Tr(W\rho)+\Tr(W\sigma)\big|
=|\Tr(W(\rho-\sigma))|
\le \|W\|_\infty\,\|\rho-\sigma\|_1
\le c\,\|\rho-\sigma\|_1 $\cite{audenaert2013comparisonsquantumstatedistinguishability}.
Define $f_W(\rho):=-\Tr(W\rho)$. Then $|f_W(\rho)-f_W(\sigma)|\le c\|\rho-\sigma\|_1$ for all
$W$ in feasible set.
Using the inequality of supremum
$|\sup_i a_i-\sup_i b_i|\le \sup_i |a_i-b_i|$ yields
$
|\cM_{nm}(\rho;\cF)-\cM_{nm}(\sigma;\cF)|
=\big|\sup_{W} f_W(\rho)-\sup_{W} f_W(\sigma)\big|
\le \sup_{W}|f_W(\rho)-f_W(\sigma)|
\le c\,\|\rho-\sigma\|_1,$
which proves the claim.
\end{proof}

\emph{(5) Dual problem:} Under a finite-dimensional truncation (for instance a Fock-space cutoff), the feasible set becomes finite dimensional and one can adapt Brand\~ao's Lagrange-duality argument for the entanglement-theoretic $E_{n:m}$ family to the present convex free set $\cF$\cite{Brand_o_2005}. Assuming the corresponding Slater-type feasibility condition, the primal and dual optima coincide and one obtains the representation
\begin{equation}\label{eq:dual_single}
\begin{split}
\cM_{nm}(\rho;\cF)
&=\inf\Bigl\{ms+nt:\ s,t\ge 0,\ \exists\,\sigma\in\cF,\ \tau,\delta\in D(\cH)\ \\ &\text{such that}\quad  
 \rho+s\tau=(1+s-t)\sigma+t\delta\Bigr\}.
\end{split}
\end{equation}
For $n\to\infty$ this reduces to a generalized-robustness-type quantity. For the representative case $n=m=1$, trace-norm duality together with a standard minimax argument yields a particularly transparent distance form,
\begin{equation}\label{eq:trace_distance_cone}
\cM_{11}(\rho;\cF)=\inf_{\tau\in \overline{\cone(\cF)}}\|\rho-\tau\|_1,
\end{equation}
where the infimum is over the trace-norm closure of the cone generated by free states. Let $C:=\overline{\cone(\cF)}$. Then:
\begin{proof}
 Using trace-norm duality \cite{Watrous_2018},
\[
\|X\|_1=\sup_{\|W\|_\infty\le 1}\Tr(WX),
\]
we write
\[
\inf_{\tau\in C}\|\rho-\tau\|_1
=
\inf_{\tau\in C}\sup_{\|W\|_\infty\le 1}\Tr[W(\tau-\rho)].
\]
A standard minimax interchange then gives
\[
\inf_{\tau\in C}\|\rho-\tau\|_1=
\sup_{\|W\|_\infty\le 1}\inf_{\tau\in C}\Tr[W(\tau-\rho)].
\]
Since $C$ is a cone, the inner infimum equals $-\Tr(W\rho)$ whenever $W\in C^*$, the dual cone of $C$, and equals $-\infty$ otherwise. Hence
\[
\inf_{\tau\in C}\|\rho-\tau\|_1=
\sup_{\substack{W\in C^*\\ \|W\|_\infty\le 1}}\bigl(-\Tr(W\rho)\bigr)=\cM_{11}(\rho;\cF),
\]
which proves \eqref{eq:trace_distance_cone}.
\end{proof}
Specializing to states where $\cF$ is $\Wplus$ or $\bar\cG_c$ we get different families of measure for WN or GNG. Specifically for the set of Wigner positive states we can define \emph{negativity depth} $\Delta(\rho):$
\begin{align}
    \Delta(\rho)=\max_{\alpha}[-W_\rho(\alpha)]_+
\end{align}
which picks out the largest negative value attained by the Wigner function. Because displaced parity operators are feasible WN witnesses after the appropriate box rescaling\cite{Chabaud_2021}, this quantity yields the lower bound
\begin{align}
        \frac{\pi}{2}\Delta(\rho)\min\{m,n\}\le \cM_{nm}(\rho;\Wplus)\le n.
    \end{align}
\subsection{Non-convex $\cF$}\label{fam3}
Let $\cF$ now be trace-norm closed but nonconvex, as in the case of Gaussian states. A single-copy linear witness $W$ that is nonnegative on $\cF$ is automatically nonnegative on $\cl(\conv(\cF))$ by linearity and continuity. Consequently, any single-copy witness monotone that vanishes on $\cF$ must also vanish on $\cl(\conv(\cF))$, and therefore cannot be faithful to $\cF$ whenever $\cl(\conv(\cF))\setminus \cF\neq\varnothing$. The key observation is that two copies allow quadratic functionals of $\rho$ to be realized as linear functionals of $\rho^{\otimes 2}$\cite{Turner_2025}. This additional expressive power restores separation for trace-norm closed nonconvex free sets.

We therefore introduce the diagonal two-copy lift
\begin{equation}
\cF^{\otimes 2}:=\{\sigma^{\otimes 2}:\sigma\in \cF\},
\qquad
\widetilde \cF^{(2)}:=\cl\!\big(\conv(\cF^{\otimes 2})\big),
\end{equation}
and define the two-copy witness box
\begin{equation}
K^{(2)}_{nm}:=\{W=W^\dagger\in B(\cH^{\otimes 2}):-n\id\le W\le m\id\}.
\end{equation}
The faithful two-copy monotone is then
\begin{equation}\label{eq:def_two_copy_Enm}
\cM^{(2)}_{nm}(\rho;\cF)
:=\sup_{W\in \Wit(\widetilde \cF^{(2)})\cap K^{(2)}_{nm}}[-\Tr(W\rho^{\otimes 2})]_+.
\end{equation}
This construction recovers faithfulness while retaining the witness-based operational interpretation. The analogue of convexity is no longer automatic on the original single-copy state space, but the remaining structural properties survive in the following form.

\emph{(1) Faithfulness:} $\cM^{(2)}_{nm}(\rho;\cF)=0$ if and only if $\rho\in\cF$.
\begin{proof}[Proof sketch]
If $\rho\in \cF$, then $\rho^{\otimes 2}\in \cF^{\otimes 2}\subseteq\widetilde \cF^{(2)}$, and hence $\Tr(W\rho^{\otimes 2})\ge 0$ for every $W\in \Wit(\widetilde \cF^{(2)})$. Therefore $\cM^{(2)}_{nm}(\rho;\cF)=0$. Conversely, if $\rho\notin\cF$ and $\cF$ is trace-norm closed, the two-copy separation result of Ref.~\cite{Turner_2025} provides a bounded Hermitian operator $W_0$ on $\cH^{\otimes 2}$ such that $\Tr(W_0\rho^{\otimes 2})<0$ while $\Tr(W_0\sigma^{\otimes 2})\ge 0$ for all $\sigma\in\cF$. By convexity and continuity this implies $W_0\in\Wit(\widetilde \cF^{(2)})$. Rescaling by $t=\min\{n,m\}/\|W_0\|_\infty$ enforces the box constraint without changing the sign, so the measure is strictly positive.
\end{proof}

\emph{(2) Strong monotonicity in the lifted theory:} Define the lifted functional
\begin{equation}
\widetilde \cM^{(2)}_{nm}(\Omega;\cF)
:=
\sup_{W\in \Wit(\widetilde\cF^{(2)})\cap K^{(2)}_{nm}}\bigl[-\Tr(W\Omega)\bigr]_+,
\end{equation}
for arbitrary $\Omega\in D(\cH^{\otimes 2})$. Let $\{\Gamma_\ell\}_\ell$ be a CP instrument on $\cH^{\otimes 2}$ such that
\begin{equation}
\Gamma_\ell(\Xi)\in \cone(\widetilde\cF^{(2)})
\qquad
\forall\,\Xi\in \widetilde\cF^{(2)},\ \forall\,\ell,
\end{equation}
and assume that $\sum_\ell \Gamma_\ell$ is trace preserving. Then for every $\Omega\in D(\cH^{\otimes 2})$,
\begin{equation}
\widetilde \cM^{(2)}_{nm}(\Omega;\cF)
\ge
\sum_\ell q_\ell\,\widetilde \cM^{(2)}_{nm}(\Omega_\ell;\cF),
\end{equation}
where
\begin{equation}
q_\ell:=\Tr[\Gamma_\ell(\Omega)],
\qquad
\Omega_\ell:=\Gamma_\ell(\Omega)/q_\ell
\quad (q_\ell>0).
\end{equation}
\begin{proof}
For each branch $\ell$, choose a feasible witness $W_\ell\in \Wit(\widetilde\cF^{(2)})\cap K^{(2)}_{nm}$ and define
\[
W:=\sum_\ell \Gamma_\ell^\dagger(W_\ell).
\]
For any $\Xi\in \widetilde\cF^{(2)}$,
\[
\Tr(W\Xi)=\sum_\ell \Tr\!\bigl(W_\ell\Gamma_\ell(\Xi)\bigr).
\]
By assumption each $\Gamma_\ell(\Xi)$ lies in $\cone(\widetilde\cF^{(2)})$, and every $W_\ell$ is nonnegative on that cone. Hence $\Tr(W\Xi)\ge 0$ for all $\Xi\in \widetilde\cF^{(2)}$, so $W\in \Wit(\widetilde\cF^{(2)})$. Moreover, positivity of $\Gamma_\ell^\dagger$ and the bounds $-n\id\le W_\ell\le m\id$ imply
\[
-n\,\Gamma_\ell^\dagger(\id)\le \Gamma_\ell^\dagger(W_\ell)\le m\,\Gamma_\ell^\dagger(\id).
\]
Summing over $\ell$ and using $\sum_\ell \Gamma_\ell^\dagger(\id)=\id$ gives $-n\id\le W\le m\id$, hence $W\in K^{(2)}_{nm}$. Finally,
\[
-\Tr(W\Omega)
=-\sum_\ell \Tr\!\bigl(W_\ell\Gamma_\ell(\Omega)\bigr)
=\sum_\ell q_\ell\bigl[-\Tr(W_\ell\Omega_\ell)\bigr].
\]
Taking the supremum over all branch witnesses on the right and using feasibility of $W$ on the left proves the claim.
\end{proof}

\emph{Deterministic monotonicity on the original single-copy theory:} If $\Lambda$ is a free CPTP map on the original resource theory, i.e. $\Lambda(\cF)\subseteq \cF$, then
\begin{equation}
\cM^{(2)}_{nm}(\Lambda(\rho);\cF)\le \cM^{(2)}_{nm}(\rho;\cF)
\qquad
\forall\rho\in D(\cH).
\end{equation}
\begin{proof}
Apply the preceding theorem to the one-outcome instrument $\Gamma:=\Lambda\otimes\Lambda$. For every $\sigma\in\cF$,
\[
\Gamma(\sigma^{\otimes 2})=\Lambda(\sigma)^{\otimes 2}\in \cF^{\otimes 2},
\]
and therefore, by linearity and trace-norm continuity, $\Gamma(\widetilde\cF^{(2)})\subseteq \widetilde\cF^{(2)}\subseteq \cone(\widetilde\cF^{(2)})$. The lifted strong-monotonicity theorem then gives
\[
\widetilde\cM^{(2)}_{nm}\bigl((\Lambda(\rho))^{\otimes 2};\cF\bigr)
\le
\widetilde\cM^{(2)}_{nm}(\rho^{\otimes 2};\cF),
\]
which is precisely the desired inequality.
\end{proof}

\paragraph*{Remark.}
The nontrivial point is that strong monotonicity does \emph{not} automatically descend from arbitrary single-copy free instruments. The obstruction is that the faithful lifted free set $\widetilde\cF^{(2)}=\cl(\conv\{\sigma^{\otimes 2}:\sigma\in\cF\})$ is generally not preserved branchwise by product instruments built from single-copy branches, since $(\Lambda_i\otimes\Lambda_j)(\sigma^{\otimes 2})=\Lambda_i(\sigma)\otimes\Lambda_j(\sigma)$ need not belong to $\widetilde\cF^{(2)}$ when $i\neq j$. Thus the correct strong-monotonicity statement lives naturally in the lifted two-copy theory, while the original single-copy functional inherits deterministic monotonicity via the diagonal lift $\Lambda\otimes\Lambda$. Again these measures are invariant under free unitary operations (similar to the case of convex measures) $U_F$ meaning $\cM^{(2)}_{nm}(\rho;\cF)=\cM_{nm}^{(2)}(U_F\rho U_F^\dagger;\cF)$ which follows because for any two copy witness $W$ in the feasible set $U_FWU_F$ also lies in the feasible set.\\

\emph{(3) Lipschitz continuity:} Let $c:=\max\{n,m\}$. Then for all states $\rho,\sigma$,
\begin{equation}\label{eq:twocopy_lipschitz}
\big|\cM^{(2)}_{nm}(\rho;\cF)-\cM^{(2)}_{nm}(\sigma;\cF)\big|
\le 2c\,\|\rho-\sigma\|_1.
\end{equation}
\begin{proof}[Proof sketch]
For any feasible $W$ we have $\|W\|_\infty\le c$, hence
\[
|\Tr[W(\rho^{\otimes 2}-\sigma^{\otimes 2})]|\le c\,\|\rho^{\otimes 2}-\sigma^{\otimes 2}\|_1.
\]
Using $|\sup_i a_i-\sup_i b_i|\le \sup_i|a_i-b_i|$ gives
\[
|\cM^{(2)}_{nm}(\rho;\cF)-\cM^{(2)}_{nm}(\sigma;\cF)|\le c\,\|\rho^{\otimes 2}-\sigma^{\otimes 2}\|_1.
\]
Finally,
\[
\rho^{\otimes 2}-\sigma^{\otimes 2}=(\rho-\sigma)\otimes \rho+\sigma\otimes(\rho-\sigma),
\]
and $\|A\otimes B\|_1=\|A\|_1\|B\|_1$ imply
\[
\|\rho^{\otimes 2}-\sigma^{\otimes 2}\|_1\le 2\|\rho-\sigma\|_1.
\]
\end{proof}

\emph{(4) Dual problem:} Under a finite-dimensional cutoff, the same finite-dimensional duality argument used in the convex case applies to the lifted convex free set $\widetilde\cF^{(2)}$. Thus, under the corresponding Slater-type feasibility condition,
\begin{equation}\label{eq:dual_two_copy}
\begin{split}
\cM^{(2)}_{nm}(\rho;\cF)
=&\inf\Bigl\{ms+nt:\ s,t\ge 0,\ \exists\,\Xi\in \widetilde\cF^{(2)},\ \tau,\delta\in D(\cH^{\otimes 2})\ \\
&\text{such that}\
\rho^{\otimes 2}+s\tau=(1+s-t)\Xi+t\delta\Bigr\}.
\end{split}
\end{equation}
In the representative case $n=m=1$ this becomes
\begin{equation}\label{eq:twocopy_distance_form}
\cM^{(2)}_{11}(\rho;\cF)=\inf_{\tau\in \overline{\cone(\widetilde \cF^{(2)})}}\big\|\rho^{\otimes 2}-\tau\big\|_1.
\end{equation}
\\
\emph{(5) Hierarchy between the monotones:} For the three CV free sets one obtains the hierarchy
\begin{align}
    \cM_{nm}(\rho; \Wplus)\le \cM_{nm}(\rho; \bar\cG_c)\le \cM^{(2)}_{nm}(\rho; \cG).
\end{align}
The first inequality follows directly from $\bar\cG_c\subset \Wplus$. For the second, every feasible single-copy witness $W\in\Wit(\bar\cG_c)\cap K_{nm}$ lifts to the two-copy witness $W\otimes \id\in\Wit(\widetilde\cG^{(2)})\cap K^{(2)}_{nm}$, since
\[
\Tr[(W\otimes \id)\sigma^{\otimes 2}]=\Tr(W\sigma)\ge 0\qquad \forall\sigma\in\cG,
\]
and therefore also on $\widetilde\cG^{(2)}$ by convexity and closure. Evaluating on $\rho^{\otimes 2}$ gives
\[
-\Tr[(W\otimes \id)\rho^{\otimes 2}]=-\Tr(W\rho),
\]
so every value achievable in the GNG optimization is also achievable in the lifted SNG optimization.
\section{Activation of resources}\label{sec:actt}
We now connect the witness monotones to discrete-variable resources, namely two-qubit entanglement and EPR steering. The activation problem is to construct a CPTP map that converts CV resource content into an experimentally accessible DV signature while remaining faithful on free inputs. The relevant channels below are \emph{witness dependent}: for each feasible witness $W$ we build a measure-and-prepare channel whose output is a Werner state. Optimizing over witnesses then yields an operational interpretation of the witness monotones.
\subsection{Entanglement activation}
Fix a feasible witness $W$ satisfying $-\id\le W\le \id$ and define the POVM elements
\begin{align}
    M_\pm=\frac{\id\pm W}{2}.
\end{align}
We consider the measure-and-prepare channel
\begin{align}
    \Phi_W(\rho)=\Tr(M_-\rho)\,\sigma_1+\Tr(M_+\rho)\,\sigma_2,
\end{align}
where $\sigma_1=\ket{\psi^-}\bra{\psi^-}$ and $\sigma_2=(\id-\ket{\psi^-}\bra{\psi^-})/3$. The output is therefore a Werner state in the $p$-parametrization with $p=\Tr(M_-\rho)$. Its negativity is
\begin{align}
        E(\Phi_W(\rho))=
        \begin{cases}
            \dfrac{-\Tr(W\rho)}{2}, & \Tr(W\rho)<0,\\[1ex]
            0, & \Tr(W\rho)\ge 0.
        \end{cases}
    \end{align}
Hence every free input $\rho\in\cF$ is mapped to a separable state, because witnesses are nonnegative on $\cF$. Whenever the witness detects the input, the output becomes entangled. Taking the supremum over all feasible witnesses yields the exact operational relation
\begin{equation}
    \sup_{W\in \Wit(\cF)\cap K_{11}} E(\Phi_W(\rho))=\frac{1}{2}\cM_{11}(\rho;\cF).
\end{equation}
If the supremum is attained, any maximizing witness $W_o$ realizes the optimal channel.

For nonconvex free sets the faithful construction is lifted to two copies:
\begin{align}
    \sup_{W\in \Wit(\widetilde \cF^{(2)})\cap K^{(2)}_{11}} E(\Phi^{(2)}_W(\rho^{\otimes 2}))=\frac{1}{2}\cM^{(2)}_{11}(\rho; \cF),
\end{align}
where
\begin{align}
    \Phi^{(2)}_W(\rho^{\otimes 2})=\Tr(M_-\rho^{\otimes 2})\,\sigma_1+\Tr(M_+\rho^{\otimes 2})\,\sigma_2.
\end{align}
Thus the lifted two-copy monotone is again exactly the optimal entanglement yield within the witness-generated activation family.

For Wigner negativity, displaced parity witnesses provide an especially transparent lower bound. Using $W_\rho(\alpha)=\frac{2}{\pi}\Tr[\Pi(\alpha)\rho]$, one finds
\begin{align}
    \sup_{\alpha}E(\Phi_{\hat\Pi(\alpha)}(\rho))=\frac{\pi}{4}\Delta(\rho),
\end{align}
so every negative phase-space region directly yields a proportional entanglement output.

The same channel also gives a useful discord-based certification. Since $D(w(q))=q^2/2$ for Werner states, free inputs obey $q\le 1/3$, and therefore
\begin{equation}
D(\Phi_W(\rho))>\frac{1}{18}
\qquad\Longrightarrow\qquad
\rho\notin\cF.
\end{equation}
The converse is not true: $D\le 1/18$ is inconclusive, because free states already generate nonzero discord under this channel. For that reason geometric discord is not a faithful output resource for the present measure-and-prepare construction.
\begin{figure}[H]
    \centering
    \includegraphics[width=1\linewidth]{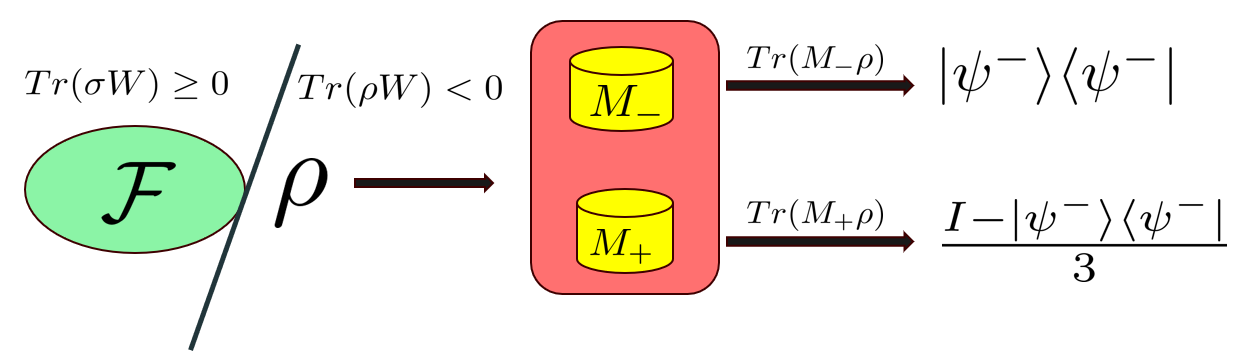}
    \caption{Witness-based activation protocol for entanglement. The POVM $\{M_+,M_-\}$ is built from the chosen witness $W$, and the measure-and-prepare channel outputs a two-qubit Werner state.}
    \label{fig:activation_protocol}
\end{figure}
\subsection{EPR steering activation}
To activate EPR steering we modify only the prepared output ensemble. We keep the same POVM but now choose $\sigma_1=\ket{\psi^-}\bra{\psi^-}$ and $\sigma_2=\id/4$, defining the channel $\bar\Phi_W$. The output is then a Werner state in the original $q$-parametrization with $q=\Tr(M_-\rho)$. Its steering monotone is
\begin{equation}
     S(\bar\Phi_W(\rho))=
     \begin{cases}
            -\Tr(W\rho), & \Tr(W\rho)<0,\\[1ex]
            0, & \Tr(W\rho)\ge 0.
        \end{cases}
\end{equation}
Again, all free states are mapped to unsteerable states, while witness-detected resourceful inputs produce steerable outputs. Optimizing over the witness family gives
\begin{equation}
    \sup_{W\in \Wit(\cF)\cap K_{11}} S(\bar\Phi_W(\rho))=\cM_{11}(\rho;\cF).
\end{equation}
The nonconvex extension is identical in spirit:
\begin{equation}
    \sup_{W\in \Wit(\widetilde \cF^{(2)})\cap K^{(2)}_{11}} S(\bar\Phi^{(2)}_W(\rho^{\otimes 2}))=\cM^{(2)}_{11}(\rho; \cF),
\end{equation}
and for displaced parity witnesses one obtains
\begin{equation}
        \sup_{\alpha}S(\bar\Phi_{\hat\Pi(\alpha)}(\rho))=\frac{\pi}{2}\Delta(\rho).
\end{equation}
Thus, within the present witness-optimized activation protocol, the bounded-witness framework provides an exact translation from CV resource content to two-qubit steering.
\section{Examples}\label{sec:examples}
We now discuss several examples that illustrate the general framework. The emphasis is on CV resources, but the witness-based construction itself is not restricted to them.
% ------------------------------------------------------------
\subsection{Maximality on odd-parity pure states }\label{subsec:odd_parity}
If $\rho$ is supported on the odd-parity subspace, equivalently $\hat\Pi\rho=-\rho$, then
\begin{equation}\label{eq:odd_maximal}
\cM_{11}(\rho; \Wplus)=\cM_{11}(\rho; \bar \cG_c)=\cM^{(2)}_{11}(\rho;\cG)=1
\end{equation}
\begin{proof}[Proof sketch]
For any state $\rho$ and any witness satisfying $-\id\le X\le \id$, one has $-\Tr(X\rho)\le 1$, so all three witness-box monotones are upper bounded by $1$. For odd-parity $\rho$, the parity operator itself is a feasible Wigner-positivity witness and satisfies $-\Tr(\Pi\rho)=1$. Hence $\cM_{11}(\rho;\Wplus)=1$. The hierarchy of monotones then gives
\[
1=\cM_{11}(\rho;\Wplus)\le \cM_{11}(\rho;\bar\cG_c)\le \cM^{(2)}_{11}(\rho;\cG)\le 1,
\]
which proves \eqref{eq:odd_maximal}.
\end{proof}
This covers many bosonic-code primitives, including odd Schr\"odinger cat states and photon-subtracted squeezed states. Operationally, such states achieve the maximal witness-box value and therefore activate the maximal entanglement and steering permitted by the present two-qubit protocol.
% ------------------------------------------------------------
\subsection{Pure-loss channel: exact evaluation and a sharp threshold}\label{subsec:loss_example}

Consider the pure-loss channel $\mathcal L_\eta$ with transmissivity $\eta\in[0,1]$. For a single photon,
\begin{equation}\label{eq:loss_single_photon}
\mathcal L_\eta(|1\rangle\langle 1|)
=\eta\,|1\rangle\langle 1|+(1-\eta)\,|0\rangle\langle 0|
=: \rho_\eta.
\end{equation}
Using the known Wigner functions of Fock states,
\(
W_{|0\rangle}(\alpha)=\frac{2}{\pi}e^{-2|\alpha|^2}
\)
and
\(
W_{|1\rangle}(\alpha)=\frac{2}{\pi}(4|\alpha|^2-1)e^{-2|\alpha|^2},
\)
we obtain the explicit radial form
\begin{equation}\label{eq:wigner_loss_onephoton}
W_{\rho_\eta}(\alpha)=\frac{2}{\pi}\,e^{-2|\alpha|^2}\Big(1-2\eta+4\eta|\alpha|^2\Big).
\end{equation}
The radial expression shows that the minimum occurs at the origin:
\begin{equation}\label{eq:wigner_min_loss}
\min_\alpha W_{\rho_\eta}(\alpha)=W_{\rho_\eta}(0)=\frac{2}{\pi}(1-2\eta).
\end{equation}
Hence Wigner negativity (and parity-activation) exhibits a sharp threshold at $\eta=\tfrac12$. For $\rho_\eta$ in~\eqref{eq:loss_single_photon},
\begin{equation}\label{eq:loss_exact_E}
\cM_{11}(\rho_\eta; \Wplus)=[2\eta-1]_+.
\end{equation}
The proof uses the following elementary boundary-mixing lemma. Let $\cF$ be any free set and consider the $n=m=1$ witness-box monotone. Assume there exist a free state $\sigma\in \cF$, a state $\tau\notin\cF$, and a feasible witness $X$ such that
\begin{equation}\label{eq:boundary_conditions}
\Tr(X\sigma)=0
\qquad\text{and}\qquad
\Tr(X\tau)=-1.
\end{equation}
Then for every $t\in[0,1]$, the mixture $\rho_t:=(1-t)\sigma+t\tau$ satisfies
\begin{equation}\label{eq:boundary_mixing_value}
\cM_{11}(\rho_t;\cF)=t.
\end{equation}

\begin{proof}[Proof sketch]
\emph{Lower bound.} Using the feasible witness $X$,
$
\cM_{11}(\rho_t;\cF)\ge -\Tr(X\rho_t)=-(1-t)\Tr(X\sigma)-t\Tr(X\tau)=t.
$
\emph{Upper bound.} For any feasible $W$ we have $\Tr(W\sigma)\ge 0$ (since $\sigma\in \cF$), and also
$\Tr(W\tau)\ge -1$ because $-I\le W\le I$ implies $\Tr(W\tau)\in[-1,1]$ for any state $\tau$.
Hence
\[
\Tr(W\rho_t)=(1-t)\Tr(W\sigma)+t\Tr(W\tau)\ge 0+t(-1)=-t,
\]
so $-\Tr(W\rho_t)\le t$ for all feasible $W$, and taking the supremum gives $\cM_{11}(\rho_t;\cF)\le t$.
Combining with the lower bound yields~\eqref{eq:boundary_mixing_value}.
\end{proof}
For $\eta\ge 1/2$ we can rewrite $\rho_\eta$ as $(1-t)\sigma+t\tau$ with $t=2\eta-1$, $\sigma=\frac{1}{2}(\ket{0}\bra{0}+\ket{1}\bra{1})\in\Wplus$, and $\tau=\ket{1}\bra{1}$. Taking the parity operator as witness, the conditions of the lemma are satisfied, and \eqref{eq:boundary_mixing_value} yields \eqref{eq:loss_exact_E}. For $\eta\le 1/2$, Eq.~\eqref{eq:wigner_min_loss} shows that the state is Wigner positive, so the monotone vanishes.
\paragraph{Interpretation.}
Equation~\eqref{eq:loss_exact_E} gives a sharp activation threshold: below $\eta=1/2$ the lossy single photon is Wigner positive and no WN-faithful activation of the present type can produce entanglement or steering; above threshold the activated entanglement grows linearly with $\eta$. 

\paragraph{Extension to general photon number states}
For an input $|n\rangle$, pure loss induces a binomial mixture of Fock states and therefore the parity expectation at the origin is
\begin{equation}\label{eq:parity_loss_fockn}
\Tr\!\big(\Pi\,\mathcal L_\eta(|n\rangle\langle n|)\big)=(1-2\eta)^n,
\end{equation}
yielding the explicit lower bound
\begin{equation}\label{eq:loss_fockn_bound}
\cM_{11}(\mathcal L_\eta(|n\rangle\langle n|);\Wplus)\ \ge\ \big[-(1-2\eta)^n\big]_+,
\end{equation}

% ------------------------------------------------------------
\subsection{Activation for GKP states}\label{subsec:gkp_example}
Among bosonic error-correcting codes, the Gottesman--Kitaev--Preskill (GKP) code\cite{gkp} is especially relevant for optical implementations because it encodes a qubit into a single oscillator while targeting small phase-space displacement errors. Since a wide class of physically relevant noise processes can be represented exactly or approximately as such displacements, GKP states are natural candidates for fault-tolerant photonic architectures\cite{Albert2018,Terhal2020}. Here we use them to illustrate how Wigner-negativity-based activation behaves as one changes the code quality.

It is useful to distinguish two related but different models.

\paragraph*{Analytic ordering via Gaussian displacement noise.}
Let $\rho_{\rm GKP}$ denote an ideal square-lattice GKP codeword. We first model finite squeezing by additive Gaussian displacement noise\cite{Noh_2019}
\begin{equation}\label{eq:gkp_noise_channel}
\begin{split}
\mathcal G_{\sigma^2}(\rho)
&:=\int_{\mathbb R^2}\frac{d^2\xi}{2\pi\sigma^2}\exp\!\Big(-\frac{\|\xi\|^2}{2\sigma^2}\Big)\,D(\xi)\rho D(\xi)^\dagger,\\
\rho_\sigma&:=\mathcal G_{\sigma^2}(\rho_{\rm GKP}),
\end{split}
\end{equation}
where $\sigma^2$ is the variance per quadrature. A common squeezing convention is
\begin{equation}\label{eq:squeezing_to_sigma}
\sigma^2=\frac12\,10^{-s/10}
\qquad (s\ \text{dB}).
\end{equation}
If $s_2>s_1$, then $\sigma_2^2<\sigma_1^2$ and the noisier state can be obtained from the less noisy one by applying an additional Gaussian channel,
\begin{equation}
    \rho_{\sigma_1}=\mathcal G_{\sigma_1^2-\sigma_2^2}(\rho_{\sigma_2}).
\end{equation}
Since Gaussian channels preserve $\Wplus$, monotonicity immediately implies that the WN witness monotones---and hence the witness-based entanglement or steering activation potential---cannot increase when the effective squeezing is reduced. This gives a clean analytic ordering statement: within this model, higher squeezing is never worse for Wigner-negativity-based activation.

\paragraph*{Finite-energy numerics with damped GKP states.}
For concrete simulations one needs normalizable finite-energy codewords. We therefore also consider the standard damped approximation
\begin{equation}
\hat N(\varepsilon)=e^{-\varepsilon \hat n},
\qquad
\hat n=\tfrac12(\hat q^2+\hat p^2-1),
\end{equation}
with approximate codewords
\begin{equation}
|\bar 0^\varepsilon\rangle \propto \hat N(\varepsilon)|\bar 0\rangle,
\qquad
|\bar 1^\varepsilon\rangle \propto \hat N(\varepsilon)|\bar 1\rangle
\end{equation}
where the damping parameter is related to the usual squeezing convention by $-10\log_{10}(\tanh\varepsilon)$\cite{Hastrup}. The analytic Gaussian-noise ordering above and the damped finite-energy model should therefore be viewed as complementary: the first supplies a rigorous monotonicity statement, while the second provides a concrete numerical testbed.

\begin{figure}[H]
    \centering
    \includegraphics[width=1\linewidth]{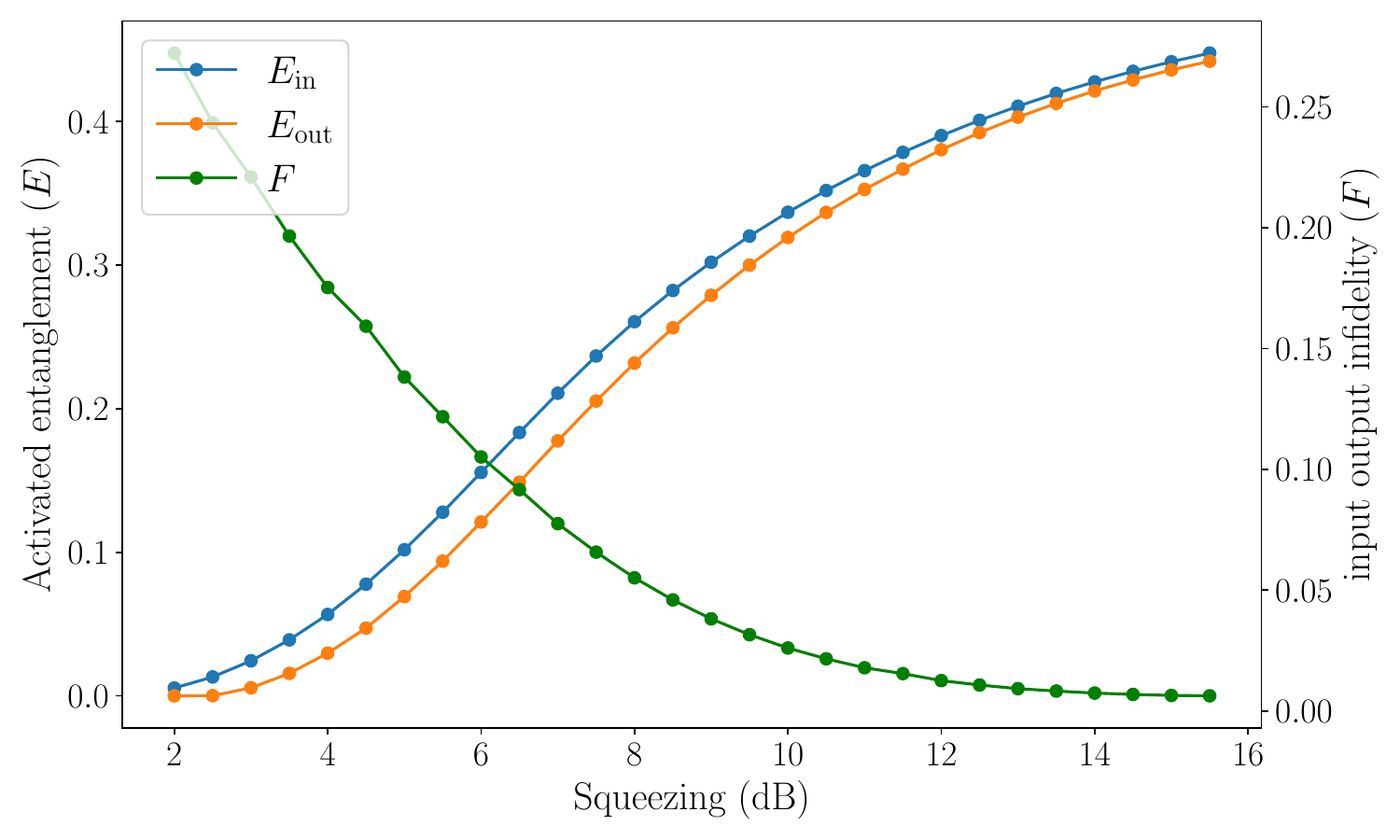}
    \caption{Finite-energy GKP example. The approximate codeword $|\bar 0^\varepsilon\rangle$ is sent through a pure-loss channel with $\eta=0.9$ and then through one round of GKP error correction. We plot the maximal entanglement activated by displaced parity witnesses for the input state ($E_{in}$) and for the final output state ($E_{out}$), together with the input--output infidelity ($F$).}
    \label{fig:placeholder}
\end{figure}

In Fig.~\ref{fig:placeholder} we take the finite-energy state $|\bar 0^\varepsilon\rangle$, send it through a pure-loss channel $\mathcal L_\eta$ with $\eta=0.9$, and then apply one round of GKP error correction\cite{Menicucci}. We plot the input--output infidelity together with the maximal entanglement activated by displaced parity witnesses, both before the loss channel and after error correction. Since displaced parity witnesses form only a distinguished witness subfamily, these curves should be interpreted as computable lower bounds on the full WN-based activation potential, not as an exact evaluation of the complete witness optimization.

The numerics indicate a clear and physically sensible trend: the pre-loss activation increases monotonically with squeezing and approaches the maximal value $0.5$ at large squeezing, while the post-correction activation remains lower throughout because residual loss and imperfect recovery reduce the surviving Wigner negativity. Thus the witness-based activation picture captures, in a directly operational way, the degradation of nonclassicality induced by loss and imperfect correction.
\subsection{Minimal activated entanglement and steering for pure states}
Evaluating GNG and SNG exactly is typically difficult because explicit optimal witnesses are rarely available. Nevertheless, for pure states one can derive useful universal lower bounds. Let $\rho_\psi:=|\psi\rangle\langle\psi|$ be a pure state and define
\[
\Lambda_\cG(\psi):=\sup_{\sigma\in \cG}\langle\psi|\sigma|\psi\rangle \in [0,1].
\]
Then
\begin{align}
\cM_{11}(\rho_\psi;\bar \cG_c) &\ge 1-\Lambda_\cG(\psi), \label{eq:GNG-lb}\\
\cM_{11}^{(2)}(\rho_\psi;\cG) &\ge 1-\Lambda_\cG(\psi)^2. \label{eq:SNG-lb}
\end{align}
\begin{proof}
For the GNG bound, consider
\[
W_\psi:=\Lambda_\cG(\psi)\,\id-\rho_\psi.
\]
For every $\omega\in \conv(\cG)$ we can write $\omega=\sum_i p_i\sigma_i$ with $\sigma_i\in\cG$, and therefore
\[
\Tr(W_\psi \omega)=\Lambda_\cG(\psi)-\sum_i p_i\langle\psi|\sigma_i|\psi\rangle\ge 0.
\]
Since $W_\psi$ is bounded, this inequality extends by trace-norm continuity to the closure $\bar\cG_c=\cl(\conv(\cG))$, so $W_\psi\in\Wit(\bar\cG_c)$. Moreover, because $0\le \Lambda_\cG(\psi)\le 1$, the spectrum of $W_\psi$ is $\{\Lambda_\cG(\psi)-1,\Lambda_\cG(\psi)\}\subset[-1,1]$, hence $-\id\le W_\psi\le \id$. Thus $W_\psi$ is feasible for $\cM_{11}(\cdot;\bar\cG_c)$ and
\[
\cM_{11}(\rho_\psi;\bar \cG_c)\ge -\Tr(W_\psi\rho_\psi)=1-\Lambda_\cG(\psi),
\]
which proves \eqref{eq:GNG-lb}.

For the SNG bound, define
\begin{equation}
W_\psi^{(2)}:=\Lambda_\cG(\psi)^2\id-\rho_\psi^{\otimes 2}.
\end{equation}
For every finite convex combination $\Xi=\sum_i p_i\sigma_i^{\otimes 2}$ with $\sigma_i\in\cG$,
\begin{align}
\Tr(W_\psi^{(2)}\Xi)
&=\Lambda_\cG(\psi)^2-\sum_i p_i\Tr(\rho_\psi^{\otimes 2}\sigma_i^{\otimes 2}) \\
&=\Lambda_\cG(\psi)^2-\sum_i p_i\langle\psi|\sigma_i|\psi\rangle^2 \ge 0.
\end{align}
Again, boundedness of $W_\psi^{(2)}$ extends this nonnegativity by continuity from $\conv(\cG^{\otimes 2})$ to its closure $\widetilde\cG^{(2)}$. Hence $W_\psi^{(2)}\in \Wit(\widetilde\cG^{(2)})$. Since $\rho_\psi^{\otimes 2}$ is a projector and $0\le \Lambda_\cG(\psi)\le 1$, the spectrum of $W_\psi^{(2)}$ is $\{\Lambda_\cG(\psi)^2-1,\Lambda_\cG(\psi)^2\}\subset[-1,1]$, so $-\id\le W_\psi^{(2)}\le \id$. Therefore $W_\psi^{(2)}$ is feasible for $\cM_{11}^{(2)}(\cdot;\cG)$, and
\begin{equation}
\cM_{11}^{(2)}(\rho_\psi;\cG)\ge -\Tr(W_\psi^{(2)}\rho_\psi^{\otimes 2})=1-\Lambda_\cG(\psi)^2,
\end{equation}
which proves \eqref{eq:SNG-lb}.
\end{proof}
In particular, if $|\psi\rangle$ is pure and non-Gaussian, then $\Lambda_\cG(\psi)<1$ because $\cG$ is closed, so both lower bounds are strictly positive. Operationally, these inequalities imply that every pure non-Gaussian state certifiably activates at least $(1-\Lambda_\cG(\psi))/2$ entanglement and $1-\Lambda_\cG(\psi)$ steering in the GNG protocol, and at least $(1-\Lambda_\cG(\psi)^2)/2$ entanglement and $1-\Lambda_\cG(\psi)^2$ steering in the lifted SNG protocol.

\section{Conclusion}
We introduced a witness-based framework for quantifying and operationally activating continuous-variable resources. For every closed convex free set $\cF$, we defined an infinite family of bounded-witness monotones $\cM_{nm}(\rho;\cF)$ and established faithfulness, strong monotonicity under free instruments, Lipschitz continuity, and convexity. For closed nonconvex free sets, we showed that a faithful single-copy linear witness theory is obstructed by convexification, and that the correct faithful formulation arises from the two-copy lift $\cM^{(2)}_{nm}(\rho;\cF)$. In that sense, the same witness-based philosophy applies uniformly to Wigner negativity, genuine non-Gaussianity, and standard non-Gaussianity, but with an intrinsically two-copy formulation in the nonconvex case.

Our main operational result is a deterministic witness-generated activation protocol that maps CV resource content into discrete-variable bipartite correlations. For the representative case $n=m=1$, we constructed measure-and-prepare channels whose outputs are two-qubit Werner states and whose optimal entanglement and EPR-steering yields are exactly proportional to the underlying witness monotones. This gives the monotones a direct operational meaning: within the present activation family, resource content is converted into experimentally accessible two-qubit nonclassical correlations.

The examples sharpen the general picture. Odd-parity states saturate the witness-box monotones and therefore yield maximal activation. For a single photon sent through a pure-loss channel, we obtained an exact formula exhibiting a sharp threshold at $\eta=1/2$ for Wigner-negativity-based activation. For GKP states, we separated a rigorous monotonicity statement based on Gaussian displacement noise from a concrete finite-energy numerical study using damped codewords under loss and one round of error correction. The resulting activation curves show that increasing squeezing enhances the witness-detected nonclassicality of the input state, while the post-correction state remains strictly less activatable because of residual loss and imperfect recovery.

Conceptually, the present work shows that witness-based quantification is not only a geometric language, but also an operational bridge from CV nonclassicality to DV quantum correlations. In the convex setting this bridge is single-copy and exact; in the nonconvex setting it remains exact after the natural two-copy lift that restores faithfulness. This suggests several directions for further work, including multimode and higher-copy constructions, sharper asymptotic analysis for finite-energy GKP families, applications to other CV or hybrid resource theories, and the systematic construction of experimentally friendly witnesses for nonconvex free sets.
\section{Acknowledgments}
We acknowledge support from the Danish National Research Foundation (bigQ, DNRF0142), EU ERC Adv (ClusterQ  grant no. 101055224) and EU Horizon Europe (CLUSTEC, grant agreement no. 101080173)
\bibliography{apssamp}

\end{document}